\documentclass[pra,floatfix,preprint,nofootinbib,eqsecnum,supersxcreiptaddeess,12pt]{revtex4}
\usepackage{epsfig}
\usepackage{bm}
\usepackage{amsmath}
\usepackage{verse}
\input epsf

\numberwithin{equation}{section}
\renewcommand{\theequation}{\arabic{section}.\arabic{equation}}
\def\etal{{\it et al.}}

\def\be{\begin{equation}}
\def\ee{\end{equation}}

\begin{document}

\vskip 3in

\title{\vskip 1in  The Conformal Rotation in Linearised Gravity\footnote{
This originally appeared in 
 {\it Quantum Field Theory and Quantum Statistics}
 (edited by  I. A. 
Batalin, C. J. Isham and G. A. Vilkovisky), Adam Hilger, Bristol, 67-87,
(1987). It has been reset and posted to arXiv to make it more accessible. An abstract has been added. An Appendix  deriving the measure for the functional integrals has been omitted.  References have not been updated nor the spelling changed.}}

\begin{abstract}
We consider the quantum mechanics of Einstein gravity linearised about flat spacetime. The two transverse-traceless components of the metric perturbation are the true physical degrees of freedom. They appear in the quantum theory as free quantum fields. Like the full Einstein  action, the Euclidean action for linearised gravity is unbounded below.  It is therefore  not possible to use that action to represent the ground state wave function as a Euclidian functional integral of the form  $\int\exp{[-(action) /\hbar]}$.  However, it is possible to represent the ground state as a Euclidian integral over the (deparametrised) action involving only the  true physical degrees of freedom.   Starting from this integral representation of the ground state and using the  techniques of  Faddeev and Popov we show how to construct a Euclidean functional integral for the ground state wave function.  The integral  explicitly exhibits the theory's gauge symmetry, locality,  and  $O(4)$ invariance. The conformal factor appears naturally rotated into the complex plane. Other representations of the ground  state  are exhibited.
\end{abstract}

\author{James B.~Hartle}  {\author{Kristin Schleich\footnote{Current address of KS: Physics Dept., Uniiversity of British Columbia, Vancouver, BC, Canada.}}

\email{hartle@physics.ucsb.edu}

\affiliation{Department of Physics, University of California,Santa Barbara, CA 93106-9530}

\date{\today}

\maketitle

\begin{verse}{\it `Cheshire Puss,' [said Alice] \dots\ `would you\\
tell me, please, which way I ought to go\\
from here?' `That depends a good deal on\\
where you want to get to, said the cat.}
\end{verse}  
\quad\quad\quad\quad\quad\quad    Lewis Carroll, {\it Alice in Wonderland} 

\vspace{.2cm}

\section{INTRODUCTION}
Functional integrals have proved to be powerful tools for the investigation
of quantum field theory. Functional integrals over Minkowski space field
configurations of the form
\begin{equation}
  \int\delta\varphi(x)\exp(iS[\varphi(x)])
	\label{eq:1.1}
\end{equation}
express concretely the sum over histories formulation of quantum
mechanics for field theory. Such integrals provide a direct route from
classical action $S[\varphi(x)]$ to quantum amplitudes in a way which is easily accessible
to formal manipulation. Functional integrals of the form
\begin{equation}
  \int\delta\varphi(x)\exp(-I[\varphi(x)])
	\label{eq:1.2}
\end{equation}
where $I[\varphi]$ is a Euclidean action and $\varphi(x)$ a Euclidean field configuration,
express ground state wavefunctions or generating functions in a way which
can be made tractable for practical computation. The work of Professor
Fradkin, whose sixtieth birthday we celebrate with this volume, provides
striking evidence for the power, richness and subtlety of functional methods
when applied to field theory.

Functional methods are particularly useful in the development of theories
with invariances, such as gauge theories or parametrised theories, because
they allow these invariances to be displayed explicitly. One expects these
methods to be especially useful in the search for a quantum theory of
gravity, which has invariances of both types. Indeed, Euclidean functional
integrals for amplitudes have been proposed as the fundamental starting
point of a quantum gravitational theory, an idea which has many novel consequences
(see, for example, Hawking 1979, 1984). A natural action for
such a theory is the Euclidean version of that for Einstein's general
relativity,
\begin{equation}
  \ell^2I[g] = -2 \int_{\partial M} d^3x h^{1/2}K - \int_M d^4 x g^{1/2}R
	\label{eq:1.3}
\end{equation}
where we use units in which $\bar h = c = 1$ and $\ell = (16\pi G)^{1/2}$ is the Planck
length. This programme immediately encounters a difficulty. The Euclidean
Einstein action is not positive definite and integrals over it of the form (\ref{eq:1.2})
will diverge (Gibbons \etal\ 1978). As Gibbons \etal\ showed, the Euclidean
functional integrals can be made convergent by an additional formal
manipulation as follows: Write the metric $g$, which is the integration
variable in a gravitational functional integral, as
\begin{equation}
  g=\Omega^2\tilde g
  \label{eq:1.4}
\end{equation}
where $\tilde g$ is a representative metric in the conformal equivalence class of $g$,
fixed, say, by the condition $R(\tilde g) = 0$. The integration over metrics $g$ can be
written as an integration over metrics $\tilde g$ which satisfy this condition and an
integration over  the conformal factor $\Omega$. If the contour of the $\Omega$ integration
is distorted to complex values, the action can be made positive definite and
the Euclidean functional integrals convergent. This is called a conformal
rotation.

There is no direct analogue of the conformal rotation in most familiar
gauge theories such as electrodynamics. The actions of these theories are
typically positive when expressed in terms of the natural Euclidean
variables. A conformal rotation is, however, needed to construct the Euclidean
functional integrals of linearised gravity very much as it is needed in the full
theory of general relativity (Gibbons and Perry 1978, Hartle 1984). In view
of this lack of analogy between Einstein gravitational theories and familiar
gauge theories, it would be helpful to have a more physically based motivation
for the Euclidean gravitational integrals in their conformally rotated
form. In this article we shall provide such motivation for linearised gravity
by deriving the conformally rotated Euclidean functional integrals from the
quantum mechanics of the theory expressed in terms of its physical degrees
of freedom.

Gauge theories are formulated in terms of redundant variables. Configurations
of the variables which differ by gauge transformations are
physically equivalent. The true physical degrees of freedom of the theory
are those which distinguish physically distinct configurations. Theories in
which time is parametrised display similar properties although there are
important differences (see, for example, Hartle and Kucha\v{r} 1984a,b).

The quantum mechanics of a theory with redundant variables is most
simply discussed in terms of its physical degrees of freedom if they can be
explicitly identified. The sums over histories for quantum amplitudes, for
example, have a simple form when expressed in terms of the physical
degrees of freedom. When so expressed they may not manifestly display all
the invariances of the theory or its locality in the redundant variables.
Quantum amplitudes, however, can also be expressed by functional integrals
over the extended space of redundant variables so as to explicitly display
invariance and locality. Such expressions are not only useful for constructing
manifestly invariant perturbation theory. They are the starting point for
the quantum mechanics of those theories with redundant variables for
which, like general relativity, the physical degrees of freedom cannot be
explicitly solved for.

The expressions for amplitudes in terms of functional integrals over the
extended variables can be derived from those over the physical degrees of
freedom by systematically adding integrals over the redundant variables
(for example, Faddeev 1969, Faddeev and Popov 1967, 1973, Fradkin and
Vilkovisky 1977, Henneaux 1985). It is through the exploration of this
connection that one arrives at the correct form and measure for the functional
integrals for gauge theories on the extended variables and makes the
connection between Hamiltonian and Lagrangian quantum mechanics. The
connection has mostly been discussed for the `Lorentzian' functional
integrals of the form (1.1) but it can also be derived for the Euclidean
functional integrals using analogous techniques. It is a natural place to look
for an understanding of the conformal rotation.

When the physical degrees of freedom can be explicitly identified, the
process of connecting functional integrals in terms of the physical degrees
of freedom with those in terms of the extended variables can be explicitly
carried out. This will be the case for linearised gravity in contrast to the full
general theory of relativity. We shall, therefore, explore the connection in
the linearised theory with an eye to understanding the conformal rotation.
The techniques for adding redundant integrations to Euclidean functional
integrals will first be developed in the context of a simple model in \S2 and
then applied to linearised gravity in \S3. There, for linearised gravity, we
shall derive the conformally rotated Euclidean functional integral for a
quantum amplitude from the functional integral for that amplitude
expressed in terms of the physical degrees of freedom.

\section{EUCLIDEAN FUNCTIONAL INTEGRALS FOR  \\\indent\indent GAUGE  AND
PARAMETRISED THEORIES}

In selecting an action to summarise the dynamics of a field theory one
frequently has in mind two goals: to find an action which (1) is a local
functional of a certain set of field variables and which (2) expresses
manifestly the invariances of the theory in terms of these variables. In
electrodynamics we seek an action which is local in the potentials $A_\mu(x)$ and
which is Lorentz invariant and gauge invariant. In gravity we might seek an
action which is a local function of the metric $g_{\alpha\beta}(x)$ and which is
invariant under the group of diffeomorphisms. Meeting both goals (1) and (2)
typically means that the action involves not only the physical degrees of
freedom---those freely specifiable on an initial value surface---but redundant
variables as well. In electrodynamics, the physical degrees of freedom are the
two transverse components of the vector potential, $A^T_i(x)$. The invariant
action also involves $A_t(x)$ and the longitudinal component $A^L_i(x)$.
In the linearised theory of gravity, the physical degrees of freedom are the
transverse-traceless parts of the metric perturbation $h^{TT}_{ij}$ while the
Einstein Lagrangian involves all the other components of the metric perturbation
$h_{\alpha\beta}$ as well. In general relativity, the action is a functional of
the metric $g_{\alpha\beta}$.
There are two physical degrees of freedom at each point on an initial value
surface although the constraints cannot be solved to exhibit them explicitly.

If one relaxes the goals of locality and invariance then there are many
different forms of the action which express the physical content of a theory.
In electrodynamics and linearised gravity, for example, one can express the
action in terms of the physical degrees of freedom at the expense of Lorentz
invariance.

How does one construct a quantum theory corresponding to a classical
theory with redundant variables? If the physical degrees of freedom can be
explicitly identified then one can proceed in two steps: (1) specify quantum
amplitudes as sums over histories expressed in terms of the physical degrees
of freedom; (2) if desired, add back into the resulting functional integral,
additional integrals over the redundant degrees of freedom so as to not
affect the value of the integral but to allow the integral to manifestly display
the original invariance and locality. When the physical degrees of freedom
cannot be explicitly identified, one can proceed formally and begin with the
form of the results of this two-step process.

In the following, we would like to illustrate this procedure with a simple
model (Hartle and Kucha\v{r} 1984b). The model is too simple to illustrate all
the issues that arise but does display some typical ones in a transparent
manner. In the succeeding section, we shall apply the techniques developed
here to the case of linearised gravity.

The configuration space of the model consists of $N$ variables $q^a(t)$, $a=1,\cdots N$
which are the physical degrees of freedom and two variables $\varphi(t)$ and
$\lambda(t)$ which represent the redundant variables. The Lagrangian is a sum
of a Lagrangian for the physical degrees of freedom $\ell(q^a, \dot{q}^a)$ and a
Lagrangian for the redundant variables $\ell^g(\varphi,\dot\varphi,\lambda)$.
For $\ell$ we take
\begin{equation}
  \ell(q^a,\dot{q}^a)=\frac{1}{2}m\delta_{ab}\dot{q}^a\dot{q}^b-V(q)
  \label{eq:2.1}
\end{equation}
and for $\ell^g$
\begin{equation}
  \ell^g(\dot\varphi,\varphi,\lambda)=\frac{1}{2}\mu(\dot\varphi-\lambda)^2.
  \label{eq:2.2}
\end{equation}
The result is a simple model of a gauge theory; $l^g$ and the total Lagrangian
are invariant under gauge transformations
\begin{align}
  \varphi(t)&\to\varphi(t)+\Lambda(t)\\ \nonumber
	\lambda(t)&\to\lambda(t)+\dot\Lambda(t).
  \label{eq:2.3}
\end{align}
Since the variable $\Lambda$ occurs in equation (\ref{eq:2.2}) without time differentiation,
there is a constraint, which is that the momentum conjugate to $\varphi$ vanishes
\begin{equation}
  \pi=\partial\ell^g/\partial\dot\varphi=0.
  \label{eq:2.4}
\end{equation}
If we did not know it already, equation (\ref{eq:2.4}) would allow us to conclude
that $\varphi$ and $\lambda$ are redundant variables and that the physical degrees of
freedom are the $q^a$.

Of course, we are not typically given gauge theories in the simple form of (\ref{eq:2.1})
plus (\ref{eq:2.2}). Rather they are expressed in terms of other variables
$Q^A=Q^A(q^a,\varphi,\lambda)$ in which some invariance is manifest. The above model,
however, displays their characteristic structure. In electrodynamics for example,
$\varphi$ corresponds to $A^L_i(x)$ and $\lambda$ corresponds to $A_t(x)$ while the $q^a$
are analogous to $A^T_i(x)$. For the purposes of our model, let us imagine that
invariance and locality have fixed the form (\ref{eq:2.1}) plus (\ref{eq:2.2}).

In the quantum theory corresponding to our simple model, states are
labelled by the physical degrees of freedom, e.g. $|q^a,t\rangle$. Amplitudes may
be constructed by sums over histories in terms of the physical degrees of
freedom in both Hamiltonian and Lagrangian form. For example, the
propagator may be expressed as
\begin{equation}
  \langle q^{\prime\prime a}t^{\prime\prime}|q^{\prime a}t'\rangle=
	\int\delta^np\delta^nq\exp\left(i\int^{t^{\prime\prime}}_{t^\prime}dt
	(p_a\dot q^a-h(q,p))\right)
  \label{eq:2.5}
\end{equation}
where $h(q, p)$ is the Hamiltonian constructed from (\ref{eq:2.1})
\begin{equation}
  h(q^a,p_a)=\frac{1}{2m}\delta^{ab}p_ap_b+V(q).
  \label{eq:2.6}
\end{equation}
The sum in (\ref{eq:2.5}) is over phase space paths which begin at $q^{\prime a}$ at $t'$ and end
at $q^{\prime\prime a}$ at $t^{\prime\prime}$. The action in the exponent is the familiar canonical one while
the measure is the usual invariant `d$p$d$q/(2\pi h)$' measure on the space of
phase space paths. One can think of the functional integral in (\ref{eq:2.5}) as being
implemented in a variety of ways---time slicing for example. Corresponding
to the different ways of `putting coordinates' on the space of functions $q(t)$
and $p(t)$ there will be different explicit forms of the `measure' for the functional
integrals. We shall not consider these in any detail in this section
although we shall supply explicit expressions in the case of linearised
gravity.\footnote{lf the reader is in any doubt, these factors were considered in detail for this model
in Hartle and Kucha\v{r} (1984b), although there is an unfortunate conflict in the use
of the notation $\delta q$ between that paper and this.}

The integrals over the momenta in (\ref{eq:2.5}) can be carried out explicitly since
the Hamiltonian is quadratic in them. This yields the Lagrangian form of
the sum over histories for the propagator
\begin{equation}
  \langle q^{\prime\prime a}t^{\prime\prime}|q^{\prime a}t'\rangle=\int\delta^n q
	\exp\left(i\int^{t^{\prime\prime}}_{t'}dt \ \ell(q^a,\dot{q}^a)\right).
  \label{eq:2.7}
\end{equation}
The transition from (2.5) to (2.7) is important because in this way the form
of the measure $\delta^nq$ is derived from Hamiltonian quantum mechanics.

Some quantum amplitudes can be conveniently expressed in terms of
Euclidean sums over histories. An example, on which we shall focus for
concreteness, is the ground state wavefunction. If one expands the left-hand
side of (\ref{eq:2.5}) or (\ref{eq:2.7}) in a complete set of energy eigenstates with energies
$E_n$ and wavefunctions $\Psi_n(q^a)$, one has, for example
\begin{equation}
  \langle q^a,0|q^{\prime a},t\rangle=\Sigma_n\Psi_n(q^{\prime a})\Psi^*_n(q^a)\exp(iE_nt).
  \label{eq:2.8}
\end{equation}
If we fix $q^{\prime a}$ to be at the minimum of $V(q)$, rotate $t\to-i\tau$, and take the
limit as $\tau\to-\infty$, the ground state will provide the dominant contribution
to the right-hand side. Carrying out the same rotations on the right-hand
sides of (2.5) and (2.7) we arrive at expressions for the ground state
wavefunction $\Psi_0(q^a)$ up to a normalisation. From (\ref{eq:2.7}) one has
\begin{equation}
  \Psi_0(q^a)=N\int\delta^nq\exp\left(-\int^0_{-\infty}d\tau\ell_E(q^a,\dot{q}^a)\right)
  \label{eq:2.9}
\end{equation}
where $N$ is a normalising constant and $\ell_E$ is the Euclidean Lagrangian
\begin{equation}
  \ell_E(q^a,\dot{q}^a)=\frac{1}{2}m\delta_{ab}\dot{q}^a\dot{q}^b+V(q).
  \label{eq:2.10}
\end{equation}
The exponent in (\ref{eq:2.9}) is minus the Euclidean action. From (\ref{eq:2.5}) we also
have
\begin{equation}
  \Psi_0(q^a)=N\int\delta^np\ \delta^nq\exp\left(-\int^0_{-\infty}d\tau(h(q,p)-ip_z\dot{q}^a)\right).
  \label{eq:2.11}
\end{equation}
(Note that the momenta are not rotated in passing from (\ref{eq:2.5}) to (\ref{eq:2.11}) and
a divergent expression would result if they were.) Equation~(\ref{eq:2.11}) is perhaps
less familiar than (\ref{eq:2.9}) but it is still useful. Equation (\ref{eq:2.9}) can be derived
from (\ref{eq:2.11}) by integrating out the momenta. Most importantly (\ref{eq:2.11}) shows
that, if the Hamiltonian of the physical degrees of freedom has a lower
bound, then the Euclidean functional integrals of the theory will converge.
This will be the case for electrodynamics and for linearised gravity. It may
also be of interest for general relativity where initial data which satisfy the
constraints, and are thus restricted to the physical degrees of freedom, have
positive energy (Schoen and Yau 1979b, Witten 1981).

By adding further integrations over the redundant variables, the functional
integrals (\ref{eq:2.5}), (\ref{eq:2.7}), (\ref{eq:2.9}) and (\ref{eq:2.11}) can be expressed as integrals
over the extended variables involving the full action. Consider for example
the functional integral for the transition amplitude (\ref{eq:2.7}). For any function
$\Phi(\varphi)$ such that $\Phi(\varphi)=0$ has a unique solution, the following identity is true
\begin{equation}
  1=\int\delta\varphi\delta\lambda\det\left[\left|\frac{\partial\Phi}{\partial\varphi}\right|\right]
	\delta[\Phi(\varphi)]\exp\left(i\int^{t^{\prime\prime}}_{t'}dt \ \ell^g(\varphi,\dot{\varphi},\lambda)\right).
  \label{eq:2.12}
\end{equation}
The identity can be verified by carrying out the integral over $\lambda$---it is a
Gaussian---and then the integral over $\varphi$ using the $\delta$ function. The term
$\det[|\partial\Phi/\partial\varphi|]$ is the product of factors which depend on $\Phi$ and are
necessary to make the integral unity. In a time slicing implementation of
(\ref{eq:2.12}) there would be one factor of $|\partial\Phi/\partial\varphi|$ for each time slice. Together,
these factors make up the familiar Faddeev-Popov determinant for the
simple gauge transformation (2.3)) and the `gauge fixing condition'
$\Phi(\varphi)=0$. To emphasise this they can be written
$\det(|\partial\Phi^\Lambda/\partial\Lambda|)=\det|\partial\Phi(\varphi+\Lambda)/\partial\Lambda)|$.
Other numerical factors necessary to make the integral
exactly unity have been absorbed into $\delta\varphi\delta\lambda$. If the identity (\ref{eq:2.12}) is
inserted in the functional integral (\ref{eq:2.7}), the following expression for the
transition amplitude results:
\begin{equation}
  \langle q^{\prime\prime a}t^{\prime\prime}|q^{\prime a}t'\rangle=\int\delta^{n+2}q
	\det\left[\left|\frac{\partial\Phi^\Lambda}{\partial\Lambda}\right|\right]
	\delta[\Phi(\varphi)]\exp(iS[q^\alpha])
  \label{eq:2.13}
\end{equation}
where we have written $q^\alpha=\{q^a,\varphi,\lambda\}$ for the extended variables and $S$ is
the total action constructed from the sum of $\ell$ and $\ell^g$. Equation~(\ref{eq:2.13}) is the
familiar form of the functional integral for the propagator in a gauge theory
and the analysis above is the familiar derivation of it (see for example
Faddeev 1969).

The repertoire of identities which can be used to create a path integral
with the action $S[q^\alpha]$ is not limited to (\ref{eq:2.12}). For example, one might have
used
\begin{equation}
  1=\int\delta\varphi\delta\lambda\delta^s[\varphi]\delta[\lambda]
	\det\left[\left|\frac{\partial\lambda^\Lambda}{\partial\Lambda}\right|\right]
	\exp\left(i\int^{t^{\prime\prime}}_{t'}dt \ \ell^g(\varphi,\dot{\varphi},\lambda)\right)
  \label{eq:2.14}
\end{equation}
where $\delta^s[\varphi]$ is a $\delta$ function enforcing the condition $\varphi=0$ only on the final
surface $t=t^{\prime\prime}$ This identity follows because the $\lambda$ integration is fixed by its
$\delta$ function and the $\varphi$ integration is a Gaussian or is fixed by the $\delta$ function
on the surface. Inserting this in (\ref{eq:2.7}) we recover a path integral of the form
(\ref{eq:2.13}) but with a different set of gauge fixing $\delta$ functions which involve both
$\varphi$ and $\lambda$. The condition $\lambda=0$ fixes the gauge freedom of (2.3) up to transformations
of the form $\varphi\to\varphi+\Lambda$ where $\Lambda$ is constant. Fixing $\varphi$ on the
surface fixes this last bit of gauge freedom.

\renewcommand{\theequation}{\arabic{section}.\arabic{equation}a}

The above model does not display the most general type of action involving
redundant variables and the identities (\ref{eq:2.12}) and (\ref{eq:2.14}) are not the most
general ways of adding integrations over such variables to functional
integrals. For example, one might want to add gauge {\it invariant} redundant
variables (we shall see an example in linearised gravity) and certainly there
are many other forms of gauge fixing. Considerable insight into the various
possibilities and the issues that they raise can be gained by studying the
theory in its Hamiltonian form and by a study of the gauge and
reparametrisation transformations on the space of extended variables.
From the Hamiltonian theory, for example, one learns that the characteristic
form (\ref{eq:2.13}) emerges naturally from (\ref{eq:2.5}) by introducing a $\delta$ function
on the extended phase space to enforce the constraints depending on
momenta, `exponentiating' that $\delta$ function via $\delta(\pi)=(2\pi)^{-1}\int d\lambda\exp(i\lambda\pi)$
(thereby introducing a further integration over the multiplier) and integrating
out the momenta. From the study of the theory on the extended space
of variables one learns that the different possibilities for introducing
redundant variables exemplified by (\ref{eq:2.12}) and (\ref{eq:2.14}) correspond to different
ways of slicing the gauge orbits on the extended space so that only physically
distinct configurations contribute to the sum over histories. We shall
not review these general insights here and indeed there is no need to do so
since they have been thoroughly discussed (Faddeev and Popov 1973,
Fradkin and Vilkovisky 1977, Hartle and Kucha\v{r} 1984a,b, Henneaux 1985
and many other references). Rather we shall only note that it is possible to
add integrations over redundant variables to the functional integrals in
terms of the physical degrees of freedom with two identities
\begin{equation}
  1=\int^{+\infty}_{-\infty}dx\delta(x)
  \label{eq:2.15a}
\end{equation}
and
\begin{equation*}\tag{2.15b}
  1=\frac{1}{\sqrt{i\pi}}\int^{+\infty}_{-\infty}dxe^{ix^2}.
  \label{eq:2.15b}
\end{equation*}
Where one goes with these identities depends on where one wants to get to.

To proceed from Euclidean functional integrals in terms of the physical
degrees of freedom to equivalent ones on an extended space of variables is
a completely analogous process to that described above. The identity
(\ref{eq:2.15a}) is still of use, but because the exponents in the Euclidean integrals
are real, (\ref{eq:2.15b}) is typically replaced by
\begin{equation*}\tag{2.15c}
  1=\frac{1}{\sqrt{\pi}}\int^{+\infty}_{-\infty}dx \ e^{-x^2}.
  \label{eq:2.15c}
\end{equation*}
As an example, consider adding integrations over $\varphi$ and $\lambda$ to the integral
(\ref{eq:2.9}) for the ground state wavefunction of our model so that the resulting
integral involves the Euclidean action for the theory. To obtain a Euclidean
version of (\ref{eq:2.2}) one may rotate $t\to-i\tau$ and also $\lambda\to i\lambda$. Thus, a Euclidean
gauge action is
\begin{equation}\tag{2.16}
  I^g=\int d\tau\frac{1}{2}\mu(\dot{\varphi}-\lambda)^2
  \label{eq:2.16}
\end{equation}
and a Euclidean action for the whole theory is
\begin{equation}\tag{2.17}
  I[q^\alpha]=\int d\tau\ell_E(q^\alpha,\dot{q}^\alpha)+I^g.
  \label{eq:2.17}
\end{equation}
The form of the Euclidean action is determined by the goals of locality and
invariance in the extended space of variables $\{q^a,\varphi,\lambda\}$
and in turn this dictates how the rotations are to be carried out. Thus, in the
above example we rotate $\lambda\to i\lambda$ and not $\lambda\to\lambda$ or
$\lambda\to-i\lambda$ so that gauge invariance in the
form (\ref{eq:2.13}) is maintained. This can be the only motivation since the
additional variables have no physical content. The process is familiar from
electrodynamics where we rotate $A_t\to iA_\tau$ as we rotate $t\to-i\tau$ to obtain
a gauge and O(4) invariant Euclidean action.

We can pass from a path integral of the form (\ref{eq:2.9}) to one involving the
action (\ref{eq:2.17}) by making use of the identity
\begin{equation}\tag{2.18}
  1=\int\delta\varphi\delta\lambda\det\left[\left|\frac{\partial\Phi^\Lambda}{\partial\Lambda}\right|\right]
	\delta[\Phi(\varphi)]\exp(-I^g[\varphi,\lambda])
  \label{eq:2.18}
\end{equation}
analogous to (\ref{eq:2.12}). It can be verified by using (\ref{eq:2.15c}) to carry out the
integrations over $\lambda$ and (\ref{eq:2.15a}) to do those over $\varphi$. Inserted in (\ref{eq:2.9}) we find
\begin{equation}\tag{2.19}
  \Psi_0[q^a]=\int\delta^{n+2}q\det\left[\left|\frac{\partial\Phi^\Lambda}{\partial\Lambda}\right|\right]
	\delta[\Phi(\varphi)]\exp(-I[q^\alpha])
  \label{eq:2.19}
\end{equation}
where $I$ is the desired form of the action (\ref{eq:2.17}).

The above procedure works when the constant $\mu$ in (\ref{eq:2.16}) is positive. It
fails when $\mu$ is negative. This can be seen either from the final answer or
from the steps through which it was derived. In the final answer, the action
$I$ is neither positive definite nor bounded below if $\mu$ is negative. In the
intermediate step, the integral (\ref{eq:2.18}) diverges.

Has the sum over histories formulation of quantum mechanics then
somehow failed for the theory (\ref{eq:2.17}) with negative $\mu$? Are Euclidean
methods inapplicable in such a theory? The answer to both questions is
certainly no. The theory in terms of the physical variables is well defined
and Euclidean methods can be applied as long as the energy is positive on
the physical degrees of freedom.

In the case of negative $\mu$ we {\it have} failed to cast the Euclidean functional
integrals of the theory into a form constructed from the action (\ref{eq:2.17}). That
action, in particular the sign of $\mu$, was assumed fixed by the requirements
of locality and invariance. There may, however, be many actions on the
extended variables which meet these requirements {\it partially}, which are
physically equivalent, and for which the corresponding Euclidean functional
integrals are convergent. For example, if we change $\mu$ to $-\mu$ in (\ref{eq:2.16})
we obtain an action which is positive definite, which is gauge invariant, and
which is physically equivalent since the gauge variables are redundant. It
only fails to meet some requirement of locality expressed in terms of
variables which mix $q^a$, $\varphi$ and $\lambda$. This action could formally be regarded
as arising from (\ref{eq:2.17}) by a further complex rotation of $\varphi$ and $\lambda$. A Euclidean
functional integral for the ground state wavefunction which involves
this new action can be derived from (\ref{eq:2.9}) because the corresponding identity
(\ref{eq:2.18}) is now convergent. Such an expression can be useful.

Starting from a quantum theory formulated in terms of physical degrees
of freedom, there are many paths leading from its Euclidean functional integrals
to those involving extended variables. How one proceeds depends
not only on where one wants to get but also on whether there is a path
leading there. The issue of whether the quantum theory is well defined,
however, depends not on the properties of the theory expressed in terms of
extended variables but rather on its properties expressed in terms of the true
physical degrees of freedom.

\section{LINEARISED GRAVITY}

The transition between Euclidean functional integrals over physical degrees
of freedom and those over extended variables can be explicitly worked out
for the linearised version of Einstein's general relativity. This is because the
physical degrees of freedom of linearised gravity can be explicitly identified
and because its action is a quadratic functional. In this section we shall
make this transition for the Euclidean integral defining the ground state
wavefunctional for linearised gravity using the techniques reviewed in \S2.

The action for linearised gravity is obtained from that of general relativity
by expanding the metric in small perturbations $h_{\alpha\beta}$ about flat space. We
shall assume throughout that these metric perturbations fall off spatially as
$1/r^{3/2}$ or better at infinity. This will be a sufficient class of perturbations for
our purposes. The action is then
\begin{equation}\tag{3.1}
  \ell^2S_2[h_{\alpha\beta}]=\frac{1}{2}\int_Md^4x(h^{\alpha\beta}G_{\alpha\beta})
	+\frac{1}{2}\int_{\partial M}d^3x\ h^{ij}(K_{ij}-\delta_{ij}K^k_k)
  \label{eq:3.1}
\end{equation}
where, in this section, $G_{\alpha\beta}$ is the {\it linearised} Einstein tensor and $K_{ij}$ is the
linearised extrinsic curvature of a constant $t$ boundary of the region of
interest. The action is invariant under gauge transformations of the form
\begin{equation}\tag{3.2}
  h_{\alpha\beta}\to h_{\alpha\beta} + \nabla_{(\alpha}\xi_{\beta)}
  \label{eq:3.2}
\end{equation}
and as a consequence the theory has four constraints. The four constraints
and the four gauge degrees of freedom mean that eight of the ten $h_{\alpha\beta}$ are
redundant variables while the remaining two are the physical degrees of
freedom of linearised gravity. These can be found by writing the theory in
$3 + 1$ form to exhibit its initial value formulation and then solving the
constraints on an initial constant $t$ slice (see Arnowitt and Deser 1959). The
familiar result is that the physical degrees of freedom are the two
transverse-traceless components of the perturbation in the metric of a constant
$t$ three-surface, $h^{TT}_{ij}$. That is, if the metric $h_{ij}$ of this surface (the spatial
components of $h_{\alpha\beta}$) is analysed into Fourier components labelled by a
wavevector $k^i$, then the two trace-free components of $h_{ij}$ projected into the
subspace transverse to $k^i$ are the physical degrees of freedom. In terms of
them, the action is\footnote{Throughout greek indices range over four dimensions while latin indices range over
three. The signature is ( - , +, +, +) when we are discussing Lorentzian
space-times and ( +, +, +, +) for Euclidean ones.}
\begin{equation}\tag{3.3}
  \ell^2S_2=\frac{1}{4}\int d^4x[(\dot{h}^{TT}_{ij})^2-(\nabla_ih^{TT}_{jk})^2)]
  \label{eq:3.3}
\end{equation}
where we have introduced the obvious convention that for any tensor
$(a_{ij}\dots)^2=a_{ij}\dots a^{ij\dots}$ and a similar one in four dimensions.
The corresponding Hamiltonian is
\begin{equation}\tag{3.4}
  \ell^2h_2=\int d^3x[(\pi^{TT}_{ij})^2+\frac{1}{4}(\nabla_ih^{TT}_{jk})^2]
  \label{eq:3.4}
\end{equation}
where $\pi^{TT}_{ij}$ is the momentum conjugate to $h^{TT}_{ij}$. We note that the Hamiltonian
is positive definite. Indeed, this is just the Hamiltonian for an
assembly of independent harmonic oscillators. The quantum theory is
therefore certainly well defined.

The ground state wavefunction for the theory (Kuchar 1970) is the
wavefunction for the state with all the oscillators in their ground states. It
can be constructed by the Euclidean functional integral analogous to (\ref{eq:2.9})
(Hartle 1984)
\begin{equation}\tag{3.5}
  \Psi_0[h^{TT}_{ij},T]=\int\delta h^{TT}_{ij}\exp(-i_2[h^{TT}_{ij}])
  \label{eq:3.5}
\end{equation}
where $i_2$ is the Euclidean action for linearised gravity and the sum is over
all transverse-traceless tensor field configurations in the half space $x^0<T$
that match the argument of the wavefunction on the surface $x^0=T$ and
which fall off fast enough at Euclidean infinity so that the action is finite.
We shall exhibit the measure in the Appendix. Explicitly, $i_2$ is
\begin{equation}\tag{3.6}
  \ell^2i_2=\frac{1}{4}\int d^4x[(\dot{h}^{TT}_{ij})^2+(\nabla_ih^{TT}_{jk})^2].
  \label{eq:3.6}
\end{equation}
It is positive definite and the integral (\ref{eq:3.5}) therefore converges. This could
be seen in a different way from the positivity of the Hamiltonian and the
analogue of (\ref{eq:2.11}).

Equation~(\ref{eq:3.5}) is where we start. We would like to add redundant integrations
to this expression until we arrive at an expression for $\Psi_0$ which is
manifestly gauge invariant and O(4) invariant. An O(4) and gauge invariant
Euclidean action which is also local in the metric perturbations is the
linearised version of (\ref{eq:1.3}),
\begin{align}\tag{3.7}
  \ell^2I_2=\frac{1}{4}&\int_Md^4x[(\nabla_\alpha\bar h_{\beta\gamma})(\nabla^\alpha h^{\beta\gamma})-2(\nabla^\alpha\bar h_{\alpha\beta})^2]\\
	&+\left(\ \parbox[c]{5.5cm}{surface terms which involve\\
	                       only the redundant variables}\right)\nonumber
  \label{eq:3.7}
\end{align}
where
\begin{equation}\tag{3.8}
  \overline{h}^\alpha_\beta=h^\alpha_\beta-\frac{1}{2}\delta^\alpha_\beta h^\gamma_\gamma.
	\label{eq:3.8}
\end{equation}
We cannot end up with a functional integral for $\Psi_0$ involving this action.
It is not positive definite. In particular on perturbations of the special form
$h_{\alpha\beta} = - 2\delta_{\alpha\beta}\chi$ we have
\begin{equation}\tag{3.9}
  \ell^2 I_2=-6\int d^4 x (\nabla_\alpha \chi)^2.
	\label{eq:3.9}
\end{equation}
However, (3.7) is not the only gauge invariant O(4) invariant action for
linearised gravity.

To add back the redundant integrations we decompose $h_{\alpha\beta}$ into pieces
corresponding to the physical degrees of freedom and pieces corresponding
to the redundant integrations. As the result (\ref{eq:3.9}) suggests, it is convenient
to begin by decomposing $h_{\alpha\beta}$ into conformal equivalence classes as
\begin{equation}\tag{3.10}
  h_{\alpha\beta}=\varphi_{\alpha\beta}+2\chi\delta_{\alpha\beta}
	\label{eq:3.10}
\end{equation}
where the decomposition can be fixed by the O(4) invariant, gauge invariant
condition
\begin{equation}\tag{3.11}
  R(\varphi)=\nabla_\alpha \nabla_{\beta} {\varphi}^{\alpha\beta}-\nabla^2\varphi^\beta_\beta=0
	\label{eq:3.11}
\end{equation}
so that $\chi$ can be defined in terms of $h_{\alpha\beta}$ through
\begin{equation}\tag{3.12}
  R(h) = -6\nabla^2\chi
  \label{eq:3.12}
\end{equation}
and the boundary conditions that $\chi$ vanish on the surface $\chi^0 = T$ and at
infinity.}

The perturbation $\varphi_{\alpha\beta}$ may be further decomposed as
\begin{equation}\tag{3.13}
  \varphi_{\alpha\beta}=t_{\alpha\beta}+\ell_{\alpha\beta}+
	\varphi^T_{\alpha\beta}+\varphi^L_{\alpha\beta}
	\label{eq:3.13}
\end{equation}
where the components are defined as follows: let $n^\alpha$ be the unit vector
orthogonal to the constant $t$ surfaces. Consider the {\it families} of tensors $t_{\alpha\beta}$,
$\ell_{\alpha\beta}$, $\varphi^T_{\alpha\beta}$ and $\varepsilon^L_{\alpha\beta}$ satisfying the following conditions:

	
	\begin{equation}\tag{3.14a}
	 \nabla^\alpha t_{\alpha\beta}=0 \qquad n^\alpha t_{\alpha\beta}=0 \qquad t_\alpha^\alpha=0. 
	 \end{equation}
	 
	 \begin{equation}\tag{3.14b}
	 \nabla^\alpha\ell_{\alpha\beta}=0 \qquad\ell^\alpha_\alpha=0\qquad\int_Md^4x \ t^{\alpha\beta}\ell_{\alpha\beta}=0 
	 \end{equation}

	\begin{equation}\tag{3.14c}
	 \nabla^\alpha\varphi^T_{\alpha\beta}=0\qquad n^\alpha\varphi^T_{\alpha\beta}=0\qquad\int_Md^4x \  t^{\alpha\beta}\varphi^T_{\alpha\beta} 
	 \end{equation}
	 
	
	\begin{equation}\tag{3.14d}
	\int_Md^4x \ t^{\alpha\beta}\varphi^L_{\alpha\beta}=0\qquad\int_M d^4x \ \ell^{\alpha\beta}\varphi^L_{\alpha\beta}=0\quad\int_Md^4x \ \varphi^{T\alpha\beta}\varphi^L_{\alpha\beta}=0 .
	\end{equation}

The orthogonality conditions are understood to hold for all tensors in the
families. Then there is a unique decomposition of $\varphi_{\alpha\beta}$ into members of these
families which we write as (\ref{eq:3.13}).  The condition (\ref{eq:3.11}) fixes $\varphi^T_{\alpha\beta}$ = 0. The
tensors $t_{\alpha\beta}$ correspond to the physical degrees of freedom. The rest are
redundant.

Under gauge transformations only $t_{\alpha\beta}$, $\ell_{\alpha\beta}$ and $\chi$ are unchanged. Since the
action (3.7) is gauge invariant it can be expressed as a Lorentz invariant
combination of these quantities. In fact it has the form
\begin{align}\tag{3.15}
  \ell^2I_2=&\frac{1}{4}\int_Md^4x [(\nabla_\alpha t_{\beta\gamma})^2+(\nabla_\alpha\ell_{\beta\gamma})^2 -24(\nabla_\alpha\chi)^2] \nonumber \\
	-&\frac{1}{4}\int_{\partial M}d^3x n^ \alpha \nabla_\alpha[2(n^\beta\ell_{\beta\gamma})^2-\frac{3}{2}(n^\beta n^\gamma\ell_{\beta\gamma})^2]. \nonumber
	\label{eq:3.15}
\end{align}

Using this decomposition of the metric we can proceed as in \S2 to add
in the redundant degrees of freedom by inserting in (\ref{eq:3.5}) identities composed
of Gaussian integrals over the gauge invariant quantities and integrals
over gauge fixing $\delta$-functions for the gauge non-invariant ones. Although
the final form is independent of the gauge fixing conditions it clarifies the
argument to use a particular one. We shall choose
\begin{equation}\tag{3.16}
  C_\alpha=\nabla^\beta\overline\varphi_{\alpha\beta}=0
	\label{eq:3.16}
\end{equation}\
which, when combined with (\ref{eq:3.11}), fixes the $\varphi^L_{\alpha\beta}$ components up to a
transformation (\ref{eq:3.2}) satisfying
\begin{equation}\tag{3.17}
  \nabla^2\xi_\beta=0.
  \label{eq:3.17}
\end{equation}
By fixing a further condition on the $\chi^0 = T$ surface this remaining gauge
freedom can be fixed. Additionally, conditions at the boundary and at infinity
are needed on the remaining redundant components of $h_{\alpha\beta}$ to define
the class of configurations over which we shall integrate. For simplicity we
will take the approach of fixing all fields on the boundary by requiring $t_{\alpha\beta}$ to
match the argument of the wave function at $\chi^0 = T$, by requiring the spatial
part $h_{ij}$ of the remaining components to vanish there\footnote{Alternatively
we could integrate over redundant variables which are not fixed on
the boundary by inserting additional gauge fixing $\delta$ functions at the boundary
surface (see for example Hartle 1984).}, and to satisfy the
gauge condition (\ref{eq:3.16}). Finally all components of $h_{\alpha\beta}$ will be required to
vanish at Euclidean infinity rapidly enough so that the action is finite. On
such configurations the surface term in the action (3.15) vanishes.


In terms of the decomposition (\ref{eq:3.13}), the action $i_2$ (3.6)) on the
physical degrees of freedom takes the form
\begin{equation}\tag{3.18}
  \ell^2i_2=\frac{1}{4}\int d^4x(\nabla_\alpha t_{\beta\gamma})^2.
	\label{eq:3.18}
\end{equation}
In the class over which we plan to integrate, the most general quadratic
action in the redundant variables which is gauge invariant and O(4) invariant
in the sense of being independent of $n^\alpha$ is
\begin{equation}\tag{3.19}
  \ell^2I^g_2=\frac{1}{4}\int d^4x(\nabla_\alpha\ell_{\beta\gamma})^2+a(\nabla_\alpha\chi)^2]
	\label{3.19}
\end{equation}
where $a$ is an arbitrary positive constant. The coefficient of the $\ell_{\beta\gamma}$ terms is
fixed by the requirement that the total action be independent of $n^\alpha$. The
coefficient of $(\nabla_\alpha\chi)^2$ is unrestricted by O(4) invariance since $\chi$ is an O(4)
scalar. The constant $a$ must be positive, however, for the action to be
positive definite.

Integrals over the redundant variables involving the action (3.19) and the
gauge fixing conditions (\ref{eq:3.16}) may be added to the Euclidean functional
integral for the ground state wave function by forming the identities

\begin{equation}\tag{3.20a}
  1=\int\delta\ell\delta\varphi^L\delta\chi\delta[C^\alpha]\det
	\left[\left|\frac{\delta C^\alpha}{\delta\xi^\beta}\right|\right]\exp(-I^g_2[\ell,\chi])
	\label{eq:3.20a}
\end{equation}
and
\begin{equation}\tag{3.20b}
  1=\int\delta\varphi^T\delta[R(\varphi]\det\left[\left|\frac{\delta R}{\delta\omega}\right|\right].
  \label{eq:3.20b}
\end{equation}
In equations (3.20) the functional integrals are over the configurations we
have specified to the past of the surface $\chi^0 = T$. The determinant in (\ref{eq:3.20a})
is the Faddeev-Popov determinant of the operator constructed by varying
the gauge fixing condition $C^\alpha$ (3.16) with respect to the gauge
parameter $\xi^\alpha$ (3.2). The determinant in equation (\ref{eq:3.20b}) is of the
operator constructed by varying the condition (\ref{eq:3.11}) which fixes the conformal
equivalence class by an infinitesimal conformal transformation
\begin{equation}\tag{3.21}
  h_{\alpha\beta}\to h_{\alpha\beta}+2\delta_{\alpha\beta}\omega.
	\label{eq:3.21}
\end{equation}
A specific measure is required in order for equations (3.20) to be true. This
will be calculated explicitly in the Appendix. of the published paper.

Inserting the identities (3.20) into the Euclidean functional integral (\ref{eq:3.5})
we arrive at the following expression for the ground state wavefunction
\begin{align}\tag{3.22}
  \Psi_0[h^{TT}_{ij},T]&=\int\delta\varphi\delta\chi\delta[C^\alpha(\varphi)[R(\varphi)]\det
	\left[\left|\frac{\delta C^\alpha}{\delta\xi^\beta}\right|\right] \nonumber \\ \nonumber
	&\times\det\left[\left|\frac{\delta R}{\delta\omega}\right|\right]\exp(-\hat I_2[\varphi,\chi]).
\end{align}
Here, $\hat I_2$ is the sum of $i_2$ and $I^g_2$
\begin{equation}\tag{3.23}
  \ell^2\hat I_2[\varphi,\chi]=\frac{1}{4}\int d^4x[(\nabla_\alpha t_{\beta\gamma})^2+
	(\nabla_\alpha\ell_{\beta\gamma})^2+a(\nabla\chi)^2]
  \label{eq:3.23}
\end{equation}
where $a$ is any positive constant. The integral in equation (3.22) is over all
{\it ten} components of $\varphi_{\alpha\beta}$ {\it and} over the `conformal factor' $\chi$ in the class of
configurations described above. The integration is thus of the form of an
integration over all gauge inequivalent metrics in a conformal equivalence
class specified by $R(\varphi) = 0$ together with an integration over conformal
factor.

The action (3.23) is gauge invariant, O(4) invariant, and, for positive $a$,
it is positive definite so that the integral in (3.22) converges. If this had been
a Lorentzian functional integral we could have recovered an integral over
the action $S_2$ (equation (\ref{eq:3.1})) by choosing $a = - 24$ and carrying out the
integral over $\chi$ using the $\delta$-function of $R$. In this Euclidean case the action
cannot be made to coincide with the action $I_2$  (3.7) because, as
(3.15) shows, this would require a negative value of $a$ and lead to a
divergent functional integral. The action $\hat I_2$ is exactly that which would be
formally obtained from $I_2$ by a rotation of the conformal factor $\chi \to i\chi$ and
setting $a=24$. The action $\hat I^2$ {\it can} be expressed in terms of the metric perturbations
$h_{\alpha\beta}$ but only in a non-local manner. From (3.11)
\begin{equation}\tag{3.24}
  \hat I_2[h]=I_2[h]-\frac{(a+24)}{144}\int d^4x R(h)\nabla^{-2}R(h).
	\label{eq:3.24}
\end{equation}
This action is physically equivalent to $I_2$, gauge invariant and O(4) invariant.
As long as $a > 0$ it is positive definite. Thus, at the expense of
locality in the metric perturbations one can construct convergent functional
integrals for linearised gravity which manifestly display the invariances of
the theory. They are in fact the conformally rotated functional integrals of
Gibbons \etal\ (1978).

\section{CONCLUSIONS}

The Euclidean action for linearised gravity is not positive definite. This does
not mean that there is not a satisfactory quantum theory of the linearised
gravitational field. Neither does it mean that there is not a sum over
histories formulation of this quantum theory or that Euclidean functional
integrals cannot be used to construct appropriate amplitudes. There is a
satisfactory quantum theory because the Hamiltonian expressed in terms of
the physical degrees of freedom is positive. As a consequence there is also
a sum over histories formulation of the theory in terms of the physical
degrees of freedom and a corresponding Euclidean functional integral
construction of the ground state wavefunction.

The non-positivity of the Euclidean action for linearised gravity does
mean that we cannot express Euclidean functional integrals in a form in
which the action is manifestly local in the metric perturbations $h_{\alpha\beta}$ and O(4)
invariant. However, one can come close. One can express the Euclidean integrals
of the theory in terms of an action which is O(4) invariant and which
contains the same number of metric variables as the usual action. It is even
local when expressed in terms of the variables $\varphi_{\alpha\beta}$ and $\chi$ used in \S3. It is
only that it is non-local when expressed in terms of the metric perturbations
themselves. This action is the linearised version of the conformally rotated
action of Gibbons \etal\ (1978). (See also Gibbons and Perry (1978).)

As its name suggests, the conformally rotated action for linearised gravity
can be obtained from the Euclidean action by a formal rotation of the
conformal factor $\chi$. In a similar way, a functional integral using the
conformally rotated action may be obtained from the corresponding integral
expressed in terms of the Euclidean action by a formal rotation of
the contour of integration of the conformal factor. This is not a very satisfactory
procedure, however, because the integral involving the Euclidean
action does not exist. Neither can one start from the Lorentzian functional
integral and perform simultaneous rotations of the conformal factor and
time to obtain a Euclidean functional integral over the conformally rotated
action. There appears to be no simple distortion of both contours such that
the functional integral remains convergent at every intermediate step. Thus
the Euclidean functional integral for linearised gravity over the conformally
rotated action is not best seen as arising from some convergent functional
integral involving the usual action through a distortion of contours\footnote{it could be so seen starting from a non-local action.}.
Rather, it is best viewed as arising from the standard process of quantising
a theory with gauge and reparametrisation invariance: (1) expressing the
theory in terms of its physical degrees of freedom; (2) then formulating the
quantum sum over histories in terms of these degrees of freedom; and (3)
finally adding back in integrations over redundant variables to manifestly
express the invariance of the theory. How one adds back in these integrations
{\it is} limited in the Euclidean sums over histories by the convergence
of the final expression but is mostly determined by what final expression one
wishes to get.

That the quantum mechanics of the linearised gravitational field is well
defined and the role of the conformal factor easy to understand is no
surprise. The theory is mathematically equivalent to two harmonic
oscillators for each mode of excitation. It is of considerable interest to see
whether this understanding can be extended to linear perturbations off a
curved background, to general relativity itself and to general relativity
interacting with matter fields. The positive energy theorems of classical
general relativity (Schoen and Yau 1979b, Witten 1981) and the closely
related positive action theorems (Schoen and Yau 1979a) give hope that this
will be possible.

\section*{Acknowledgements}

One of us (JH) has benefited over the years from many discussions with
K. Kucha\v{r} on the connection between theories formulated in terms of physical
degrees of freedom and in terms of redundant variables. This work was
supported in part by NSF grant PHY 81-07384. One of us (KS) was also
supported by a Bell Laboratories Graduate Research Fellowship.
Thanks are due to Debbie Ceder for retyping the paper.

\eject

\section*{REFERENCES}

Arnowitt R and Deser S 1959 Phys. Rev. 113 745.

Faddeev L 1969 Teor. Mat. Fiz. 1 3 (Engl. transl. 1970 Theor. Math. Phys. 1 1).

Faddeev L and Popov V 1967 Phys. Lett. 25B 30

--- 1973 Usp. Fiz. Nauk 111 427 (Engl. transl. 1974 Sov. Phys. Usp. 16 777).

Fradkin E and Vilkovisky G 1977 Quantization of relativistic systems with constraints:
equivalence of canonical and covariant formalisms in the quantum
theory of the gravitational field  CERN Report TH-2332.

Gibbons G, Hawking S W and Perry M 1978 Nucl. Phys. B 138 141.

Gibbons G and Perry M 1978 Nucl. Phys. B 146 90.

Hartle J B 1984 Phys. Rev. D29 2730.

Hartle J B and Kuchar K 1984a J. Math. Phys. 25 5.

---1984b in Quantum Theory of Gravity ed S Christensen (Bristol: Adam Hilger)

Hawking S W 1979 in General Relativity: an Einstein Centenary Survey \\ 
ed by S W Hawking and W Israel (CUP).

S.W. Hawking 1984 Nucl. Phys. B 244 135.

Henneaux M 1985 Phys. Rep. 126 1.

Kuchar K 1970 J. Math. Phys. 11 3322.

Schoen R and Yau S T 1979a Phys. Rev. Lett. 42 547.

--- 1979b Phys. Rev. Lett. 43 1457.

Witten E 1981 Commun. Math. Phys. 80 381.

\end{document}